\documentclass[12pt]{article}
\usepackage{curves}
\newcommand{\be}{\begin{equation}}
\newcommand{\ee}{\end{equation}\noindent}
\newcommand{\bear}{\begin{eqnarray}}
\newcommand{\ear}{\end{eqnarray}\noindent}
\newcommand{\no}{\noindent}
\newcommand{\non}{\nonumber}

\makeatletter

\def\@authoraddress{}
\def\@title{}
\def\title#1{\gdef\@title{{\par\vskip-10pt\Large\bf
\baselineskip20pt\centering\ignorespaces\uppercase{#1}\vskip6pt}}%
\setcounter{table}{0}      \setcounter{figure}{0}
\setcounter{equation}{0}   \setcounter{section}{0}
\setcounter{subsection}{0} \setcounter{subsubsection}{0}
\setcounter{paragraph}{0}
}

\def\authors#1{\expandafter\def\expandafter\@authoraddress\expandafter
{\@authoraddress %
{\dimen0=-\prevdepth \advance\dimen0 by1.5\baselineskip
\nointerlineskip \centering
\vrule height\dimen0 width0pt\relax\ignorespaces\large\sc#1\par
}%
}%
}

\def\addresses#1{\expandafter\def\expandafter\@authoraddress\expandafter
{\@authoraddress{\nointerlineskip\vskip1pc
                 \footnotesize\it\centering\ignorespaces#1\par}}}

\def\@maketitle{%
\@title
\ifdim\prevdepth=-1000pt \prevdepth0pt\fi
\@authoraddress
}

\def\maketitle{\par
\begingroup
\let\cite\@bylinecite
\global\@topnum\z@ %
\@maketitle
\endgroup
\def\@thanks{}\def\@authoraddress{}\def\@title{}
}

\def\abstract{\par
\bgroup
\ifdim\prevdepth=-1000pt \prevdepth0pt\fi
\hsize\columnwidth
\leftskip=2em \rightskip\leftskip
\dimen0=-\prevdepth \advance\dimen0 by2pc \nointerlineskip
\noindent\vskip1.5\baselineskip\nointerlineskip\noindent\footnotesize\relax}

\newif\if@firststuff

\def\endabstract{\par
\nointerlineskip \vskip0pt
\noindent \par
\egroup
\hrule depth0pt width0pt
\global\everypar{\global\@firststufffalse}\global\@firststufftrue
}

\renewcommand\section{\@startsection {section}{1}{\z@}%
                                   {-3.5ex \@plus -1ex \@minus -.2ex}%
                                   {2.3ex \@plus.2ex}%
                                   {\normalfont\large\bfseries}}
\renewcommand\subsection{\@startsection{subsection}{2}{\z@}%
                                     {-3.25ex\@plus -1ex \@minus -.2ex}%
                                     {1.5ex \@plus .2ex}%
                                     {\normalfont\large\bfseries}}

\def\1ad{\mbox{\normalsize $^1$}}
\def\2ad{\mbox{\normalsize $^2$}}
\def\3ad{\mbox{\normalsize $^3$}}
\def\4ad{\mbox{\normalsize $^4$}}
\def\5ad{\mbox{\normalsize $^5$}}
\def\6ad{\mbox{\normalsize $^6$}}
\def\7ad{\mbox{\normalsize $^7$}}
\def\8ad{\mbox{\normalsize $^8$}}

\makeatother

\begin{document}
\input{feynman}

%%%%%%%%%%%%%%%%%%%%%%%%%%%%%%%%%%%%%%%%%%%%%%%%%%%%%%%%%%%%%%%%%%%%%%%%%%
%%%  The TITLE of the paper

\title{Four - Point Functions of Chiral Primary Operators in
$N=4$ SYM}

%%%%%%%%%%%%%%%%%%%%%%%%%%%%%%%%%%%%%%%%%%%%%%%%%%%%%%%%%%%%%%%%%%%%%%%%%%
%%%  The AUTHORS.  If some of the authors belong to different
%%%  institutions, the addresses of those institutions (which appear below)
%%%  are referred from here by commands \1ad, \2ad, etc., or the command \adref{..}

\authors{B. Eden, C. Schubert and E. Sokatchev}

%%%%%%%%%%%%%%%%%%%%%%%%%%%%%%%%%%%%%%%%%%%%%%%%%%%%%%%%%%%%%%%%%%%%%%%%%%
%%%  The ADDRESSES.  If there is more than one address, they are prepended
%%%  after the \nextaddress command

\topmargin = -.8in

\addresses{\it\small Laboratoire d'Annecy-le-Vieux de Physique
Th{\'e}orique\footnote{UMR 5108 associ{\'e}e {\`a}
 l'Universit{\'e} de Savoie} LAPTH, Chemin de Bellevue, B.P. 110,
F-74941 Annecy-le-Vieux, France \\ 
$\phantom{a}$\\
{\rm Talk given by C. Schubert at ``Quantization, Gauge Theory and
Strings'', dedicated to the memory of E.S. Fradkin, Moscow, June 5-10,
2000}
}

%%%%%%%%%%%%%%%%%%%%%%%%%%%%%%%%%%%%%%%%%%%%%%%%%%%%%%%%%%%%%%%%%%%%%%%%%%
%%% The mandatory command \maketitle actually typesets the title, author
%%% names and affiliations using the definitions above.

\maketitle

%%%%%%%%%%%%%%%%%%%%%%%%%%%%%%%%%%%%%%%%%%%%%%%%%%%%%%%%%%%%%%%%%%%%%%%%%%
%%%  The abstract shouldn't contain more than 20 lines when printed.

\begin{abstract}
We discuss recent progress in the determination of
correlators of chiral primary operators in $N=4$
Super-Yang-Mills theory, based on a combination of 
superconformal covariance arguments in $N=2$ harmonic
superspace, and Intriligator's insertion formula. Applying 
this technique to the calculation of
the supercurrent four - point function
we obtain a compact and
explicit result for its three-loop contribution
with comparatively little effort.
\end{abstract}
\vspace{15pt}
\hspace{270pt} {\rm LAPTH-Conf-814/99}
\vspace{15pt}
%%%%%%%%%%%%%%%%%%%%%%%%%%%%%%%%%%%%%%%%%%%%%%%%%%%%%%%%%%%%%%%%%%%%%%%%%%
%%%  The body of the document begins here.
%%%

As is well-known, unbroken conformal invariance imposes strong constraints
on correlators in quantum field theory.
More recently, the analysis of these constraints
has been extended to the case of superconformal invariance in
Super-Yang-Mills theories by P. Howe and P. West
\cite{1}. The natural objects to consider
in this context are finite correlators of gauge invariant composite
operators, such as the $N=4$ supercurrent.
An independent reason for the study of the same type
of correlators has been provided by Maldacena's conjecture
which relates them, at leading order in the 
${1\over N_c}$ expansion and in the strong coupling limit,
to tree-level correlators in AdS supergravity \cite{2}. 
Yet another motivation for their computation
comes from 
the operator product expansion
(see, e.g., A. Petkou's contribution to the
present proceedings).

A surprising result of those recent investigations has
been the discovery that large classes of such correlators
exist which are non-renormalized, i.e. do not receive
perturbative corrections at all \cite{3}.
Here we consider the simplest such correlator which
{\sl does} have radiative corrections, and 
explicitly compute it at the
two- and three- loop level.

\vfill\eject
Since no off-shell superspace formulation 
of $N=4$ SYM theory is known, in our loop calculations we use
a reformulation in terms of $N=2$ harmonic superfields
\cite{4}. Those live on the 
analytic 
superspace 
with coordinates 
$x^{\alpha\dot\alpha}_A,\theta^{+\alpha}, 
\bar\theta^{+\dot\alpha},u^\pm_i$. Here $u^\pm_i$ are the harmonic 
variables which form a matrix of $SU(2)$ and parametrise the 
sphere $S^2\sim SU(2)/U(1)$. A harmonic function $F^{(q)}(u^\pm)$ 
of $U(1)$ charge $q$ is a function of $u^\pm_i$ invariant under 
the action of the group $SU(2)$ (which rotates the index $i$ of 
$u^\pm_i$) and homogeneous of degree $q$ under the action of the 
group $U(1)$ (which rotates the index $\pm$ of $u^\pm_i$).

The two $N=2$ ingredients of the $N=4$ SYM theory are the $N=2$ 
SYM multiplet and the $N=2$ matter (hyper)multiplet. 
The hypermultiplet is described by an analytic 
superfield of charge $1$, $q^+(x_A,\theta^+,\bar\theta^+,u)$.
Its equation of motion is 

\begin{equation}\label{EMo}
  D^{++}q^+ = 0
\end{equation}
where $D^{++}$ is the harmonic derivative on $S^2$ (the raising 
operator of the group $SU(2)$ realised on the $U(1)$ charges, 
$D^{++}u^+=0,\; D^{++}u^-=u^+$). 
The equation of motion comes from a Dirac-like action

\begin{equation}\label{HMcov}
S_{\mbox{\scriptsize HM}} = -\int 
dud^4x_Ad^2\theta^+d^2\bar\theta^+\; \tilde q^{+}D^{++}q^+
\end{equation}
where ``$\tilde{\phantom q}$'' is an appropriate conjugation. 
By covariantising this action 
with respect to a Yang-Mills group with analytic 
parameters one introduces the
SYM gauge potential $V^{++}$, a charge 2 superfield.
Its action can be written as

\begin{equation}\label{SYMact}
  S_{\mbox{\scriptsize N=2 SYM}} =
{1\over 4g^2}\int d^4x_Ld^4\theta\;  {\rm tr}\;W^2\;. 
\end{equation}
where $W(x_L,\theta^{i\alpha})$ is the field strength tensor.
Unlike the analytic gauge potential, this is a (left-handed) chiral 
superfield which is harmonic-independent. 
Its expansion in the gauge potential is an infinite series
containing arbitrary powers of
$V^{++}$. 

When the hypermultiplet matter is taken in the adjoint 
representation of the gauge group, the sum of the 
two actions (\ref{HMcov}) 
and (\ref{SYMact}) describes the $N=4$ SYM theory, 
\begin{equation}\label{N4sym}
  S_{\mbox{\scriptsize N=4 SYM}} =  S_{\mbox{\scriptsize N=2 SYM}} +
 S_{\mbox{\scriptsize HM/SYM}}\;.
\end{equation}

The Feynman rules derived from this action are formally 
QCD - like, except that there is an infinite set of
gauge self-interactions. In our following three-loop calculation
those turn out not to contribute, so that we will have, in fact,
only QED - like Feynman diagrams.

\noindent
We consider the following $N=2$ correlator

\begin{equation}\label{1st1}
  G = \langle 
{\rm tr}\widetilde q^2_1
{\rm tr}q^2_2
{\rm tr}\widetilde q^2_3
{\rm tr}q^2_4
\rangle
\end{equation}
at the lowest Grassmann level, i.e. with 
$\bar\theta^+_{1,2,3,4}=
0$.
In $N=4$ SYM this correlator 
carries the full information on the 
correlator of four $N=4$ supercurrents \cite{5}
\footnote{Note added: 
A rigorous proof of this fact has been given only after
the conference \cite{6}.}
.
At the free field (= one-loop) level, it has a trivial
contribution

\bear
G^{\rm 1-loop} &=& \frac{16 (N_c^2-1)}
{(2\pi)^8}
{(12)(23)(34)(41)\over x_{12}^2x_{23}^2x_{34}^2x_{41}^2}.
\ear
where 
$x_{ij}\equiv x_i-x_j$ and 
$(kl)\equiv u^{+i}_ku^+_{li}\;$.

\begin{figure}[ht]
\begin{center}
\begin{picture}(42000,9000)(0,-1500)

\drawarrow[\E\ATTIP](2700,6340)
\drawarrow[\W\ATTIP](6000,0)
\drawarrow[\S\ATTIP](0,3300)
\drawarrow[\N\ATTIP](10000,3300)
\drawline\fermion[\E\REG](0,0)[10000] \global\advance\pmidx by 
-400 \global\Yone=-1500 \put(\pmidx,\Yone){a} 
\global\advance\pmidx by 400 
\drawline\gluon[\N\CENTRAL](\pmidx,\pmidy)[6] 
%\put(\pmidx,\pmidy){\circle*{1000}} 
\global\Xone=\gluonlengthy 
\drawline\fermion[\W\REG](\gluonbackx,\gluonbacky)[5000] 
\drawline\fermion[\E\REG](\gluonbackx,\gluonbacky)[5000] 
\drawline\fermion[\S\REG](\pbackx,\pbacky)[\Xone] 
\drawline\fermion[\N\REG](0,0)[\Xone] 

\drawarrow[\E\ATTIP](15700,6340)
\drawarrow[\N\ATTIP](23000,3300)
\drawarrow[\S\ATTIP](13000,3300)
\drawarrow[\W\ATTIP](19000,0)
\drawline\fermion[\E\REG](13000,0)[10000] \global\advance\pmidx by 
-400 \put(\pmidx,\Yone){b} 
\drawline\fermion[\N\REG](\pbackx,\pbacky)[\Xone] 
\drawline\fermion[\W\REG](\pbackx,\pbacky)[10000] 
\drawline\fermion[\S\REG](\pbackx,\pbacky)[\Xone] 
\global\Xtwo=\pmidx \global\Ytwo=\pfronty \startphantom 
\drawline\gluon[\NE\FLIPPED](\pmidx,\pmidy)[3] \stopphantom 
\global\Ythree=\gluonlengthy \global\negate\Ythree 
\global\advance\Ytwo by \Ythree 
\drawline\gluon[\NE\FLIPPED](\Xtwo,\Ytwo)[3] 
%\put(\pmidx,\pmidy){\circle*{1000}} 

\drawarrow[\E\ATTIP](28000,6340)
\drawarrow[\N\ATTIP](36000,3300)
\drawarrow[\S\ATTIP](26000,3300)
\drawarrow[\W\ATTIP](32000,0)
\startphantom \drawloop\gluon[\N 5](26000,0) \stopphantom 
\global\Xfive=\loopfrontx \global\negate\Xfive 
\global\advance\Xfive by \loopbackx \global\advance\Xfive by 
-10000 \global\divide\Xfive by 2 
\drawline\fermion[\E\REG](26000,0)[10000] \global\advance\pmidx by 
-400 \put(\pmidx,\Yone){c} 
\drawline\fermion[\N\REG](\pbackx,\pbacky)[\Xone] 
\drawline\fermion[\W\REG](\pbackx,\pbacky)[10000] 
\global\advance\pmidy by -2650 
%\put(\pmidx,\pmidy){\circle*{1000}} 
\global\advance\pfrontx by \Xfive 
\drawloop\gluon[\S5](\pfrontx,\pfronty) 
\drawline\fermion[\N\REG](26000,0)[\Xone] 
\end{picture}
\end{center}
\caption{Two-loop graphs}
\label{2loopgraphs}
\end{figure}
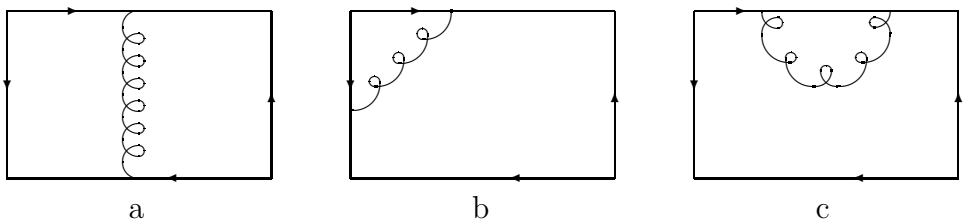

At the two-loop ($O(g^2)$) level the 
connected graphs are, up to permutations,
the ones shown in figure 
\ref{2loopgraphs}. 

\noindent
Intriligator's insertion formula \cite{7} allows one to rewrite
the sum of graphs as 

\begin{eqnarray}
\sum G_i &=& 
-{i\over 4}\int dx_5d^4\theta_5\sum G_i^{\rm ins}
\end{eqnarray}

where $G_i^{\rm ins}$ is a five-point graph obtained from
$G_i$ by inserting, into the gauge propagator, a 
${{\rm tr}W^2\over g^2}$.
The 
five-point function in the integrand is strongly constrained by
superconformal covariance, which fixes
its complete dependence on the odd variables, and
forces it to be of the form $\xi^4F(x,u)$, where
$\xi^4$ is the uniquely determined nilpotent superconformal 
five-point covariant with the required properties. 
Defining
$\rho_a^{\dot\alpha}
\equiv
(\theta_5^iu_i^+-\theta_a^+)_{\alpha}
{x^{\alpha\dot\alpha}_{5a}\over x_{5a}^2}$,
this covariant can be written as

\bear
\xi^4 &=& (34)^2\rho_1^2\rho_2^2
+2(24)(43)\rho_1^2(\rho_2\rho_3) 
+ {4\over 3}
\Bigl[(23)(41)+(12)(34)\Bigr]
(\rho_1\rho_3)(\rho_2\rho_4) 
\non\\&&
+\quad {\rm permutations}
\label{defxi}
\ear
Moreover, after the insertion our graphs factorize into
``building blocks'' which can be easily computed, and yield 
expressions which are {\sl rational in the spacetime variables}
\cite{8}.

\begin{figure}[ht]
\begin{center}
  \begin{picture}(0,4000)

  \drawline\gluon[\S\CENTRAL](0,0)[4]
  \put(\gluonbackx,\gluonbacky){\circle*{600}}
  \drawline\fermion[\W\REG](\gluonfrontx,\gluonfronty)[5000]
  \drawarrow[\E\ATTIP](\pmidx,\pmidy)
  \global\advance\pbackx by -1200
  \put(\pbackx,\pbacky) {1a}
  \drawline\fermion[\E\REG](\gluonfrontx,\gluonfronty)[5000]
  \drawarrow[\E\ATBASE](\pmidx,\pmidy)
  \global\advance\pbackx by 500
  \put(\pbackx,\pbacky) {2c}
  \global\advance\gluonbackx by 1000
  \global\advance\gluonbacky by -500
  \put(\gluonbackx,\gluonbacky) {5b}
  \global\advance\gluonfronty by 1000
    \end{picture}
  \end{center}
\vspace{45pt}
\caption{\label{blockI} Building block ``I''}
\end{figure}
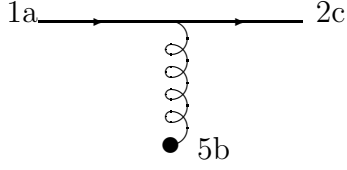
\noindent
For example, the building block shown in fig. \ref{blockI}
can be written as

\bear
I
&=&  {2gf_{abc}\over (2\pi)^4}
{(21^-)\rho_1^2+(12^-)\rho_2^2 -2(\rho_1\rho_2)\over x_{12}^2}
\non\\
\label{resultblock}
\ear\no
Summing up all graphs one reproduces the known result \cite{9},

\bear
G^{\rm 2-loop}&=& 
-32{ig^2(N_c^2-1)N_c\over (2\pi)^{12}}
{R'\over x_{12}^2x_{23}^2x_{34}^2x_{41}^2}
h^{(1)}(x_1,x_2,x_3,x_4)
\label{2loopprefinal}
\ear
where 

\bear
R'= -{1\over 2} (12)^2(34)^2
\Bigl[x_{12}^2x_{34}^2-x_{13}^2x_{24}^2-x_{14}^2x_{23}^2\Bigr]
+{\rm permutations}
\ear
The remaining integral over the insertion point is the well-known
one-loop box integral 
$h^{(1)}\equiv
\int dx_5
{1\over x_{51}^2x_{52}^2x_{53}^2x_{54}^2}$,
which can be expressed in terms of 
logarithms and dilogarithms of the conformal cross ratios.

Proceeding to the three-loop level, a repeated application of the
insertion formula allows us to write

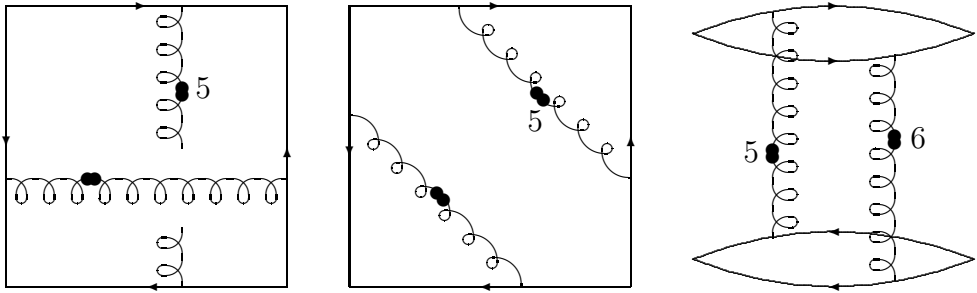
\begin{figure}[ht]
\begin{center}
\begin{picture}(42000,13000)(0,-1400)

\drawline\gluon[\E\REG](0,4100)[10] \global\Xone = \gluonbackx 
\drawline\gluon[\S\REG](6660,\Xone)[5] 
\drawline\gluon[\N\FLIPPED](6660,0)[2] 
\drawline\fermion[\E\REG](0,0)[\Xone] 
\drawarrow[\W\ATTIP](\pmidx,\pmidy) 
\drawline\fermion[\N\REG](0,0)[\Xone] 
\drawarrow[\S\ATTIP](\pmidx,\pmidy) 
\drawline\fermion[\W\REG](\Xone,\Xone)[\Xone] 
\drawarrow[\E\ATTIP](\pmidx,\pmidy) 
\drawline\fermion[\S\REG](\Xone,\Xone)[\Xone] 
\drawarrow[\N\ATTIP](\pmidx,\pmidy) \put(6660,7540){\circle*{500}} 
\put(6660,7270){\circle*{500}} \put(7160,7170){5} 
\put(3350,4100){\circle*{500}} \put(3080,4100){\circle*{500}} 
%\put(3060,4680){6} \put(5000,-3000){A} 

\drawline\fermion[\E\REG](13000,0)[\Xone] \global\advance\pmidx by 
-400 \drawarrow[\W\ATTIP](\pmidx,\pmidy) 
\drawline\fermion[\N\REG](13000,0)[\Xone] \global\advance\pmidy by 
-400 \drawarrow[\S\ATTIP](\pmidx,\pmidy) \global\Xtwo = 13000 
\global\advance\Xtwo by \Xone 
\drawline\fermion[\W\REG](\Xtwo,\Xone)[\Xone] 
\global\advance\pmidx by 400 \drawarrow[\E\ATTIP](\pmidx,\pmidy) 
\drawline\fermion[\S\REG](\Xtwo,\Xone)[\Xone] 
\global\advance\pmidy by 400 \drawarrow[\N\ATTIP](\pmidx,\pmidy) 
\drawline\gluon[\SE\FLIPPED](17160,\Xone)[6] \global\Ythree = 
\Xone \global\advance\Ythree by -3300 \global\Xthree = \Xone 
\global\advance\Xthree by 13000 \global\advance\Xthree by -3300 
\global\advance\Xthree by -250 
\put(\Xthree,\Ythree){\circle*{500}} \global\advance\Xthree by 250 
\global\advance\Ythree by -250 
\put(\Xthree,\Ythree){\circle*{500}} \global\advance\Xthree by 
-600 \put(\Xthree,5900){5} \drawline\gluon[\SE\REG](13000,6500)[6] 
\global\Xthree = 16300 \global\Ythree = 3300 
\global\advance\Xthree by 250 \put(\Xthree,\Ythree){\circle*{500}} 
\global\advance\Xthree by -250 \global\advance\Ythree by 250 
\put(\Xthree,\Ythree){\circle*{500}} \global\advance\Xthree by 120 
%\put(\Xthree,4100){6} \put(18000,-3000){B} 

\startphantom \drawline\fermion[\E\REG](26000,0)[\Xone] 
\stopphantom \global\Xfour = \pmidx \global\Xfive = \pbackx 
\global\Yfour = 1045 \global\Yfive = \Yfour \global\advance\Yfive 
by \Yfour \curve(26000,\Yfour,\Xfour,0,\Xfive,\Yfour) 
\curve(26000,\Yfour,\Xfour,\Yfive,\Xfive,\Yfour) 
\global\advance\Xfour by -200 \drawarrow[\W\ATTIP](\Xfour,0) 
\drawarrow[\W\ATTIP](\Xfour,\Yfive) \global\advance\Xfour by 200 
\global\Ysix = \Xone \global\advance\Ysix by -\Yfour 
\global\Yseven = \Ysix \global\advance\Yseven by -\Yfour 
\curve(26000,\Ysix,\Xfour,\Xone,\Xfive,\Ysix) 
\curve(26000,\Ysix,\Xfour,\Yseven,\Xfive,\Ysix) 
\global\advance\Xfour by 200 \drawarrow[\E\ATTIP](\Xfour,\Xone) 
\drawarrow[\E\ATTIP](\Xfour,\Yseven) \global\advance\Xfour by -200 
\global\Yone = 220 \global\Yeight = \Xone \global\advance\Yeight 
by -\Yone \drawline\gluon[\S\FLIPPED](29000,\Yeight)[8] 
\global\advance\Xfour by -26000 \global\advance\Xfour by -270 
\global\advance\Xfour by -135 \global\advance\Xfour by -200 
\put(27900,\Xfour){5} \global\advance\Xfour by 200 
\put(29000,\Xfour){\circle*{500}} \global\advance\Xfour by 270 
\put(29000,\Xfour){\circle*{500}} \global\advance\Xfive by -3000 
\drawline\gluon[\N\FLIPPED](\Xfive,\Yone)[8] \global\advance\Xfour 
by 540 \put(\Xfive,\Xfour){\circle*{500}} \global\advance\Xfour by 
-270 \put(\Xfive,\Xfour){\circle*{500}} \global\advance\Xfive by 
600 \global\advance\Xfour by -200 \put(\Xfive,\Xfour){6} 
%\put(31000,-3000){C} 

\end{picture}
\end{center}
\caption{Three-loop graphs}
\label{3loopgraphs}
\end{figure}

\bear G^{\rm 3-loop} &=& 
-{1\over 32}
\int d^4x_5 d^4\theta_5 \int d^4x_6 d^4\theta_6 
\Bigl\langle 
{\rm tr}\widetilde q^2_1
{\rm tr}q^2_2
{\rm tr}\widetilde q^2_3
{\rm tr}q^2_4
{\rm tr} {W_{5}^2\over g^2} 
{\rm tr} {W_{6}^2\over g^2}
\Bigr\rangle \non\\ \label{intriligation3loop} \ear\no 
Similarly to the two-loop case, the superconformal covariance of
the integrand six-point function turns out to require it to
be of the form $\xi^4\psi^4F(x,u)$,
where $\xi^4$ is our covariant above, and $\psi^4$
the corresponding covariant referring to point 6. 
This factorization is very useful since it means that,
instead of computing the component $\sim \theta_5^4\theta_6^4$
of the six-point correlator, 
which is the one that actually saturates
the integrals $\int d^4\theta_5 \int d^4\theta_6$, we can easily infer
this component from any other one. As it turns out, the component
which is the easiest one to compute in the explicit graph calculation
is the ``opposite'' one defined 
by instead setting the chiral Grassmann variables to zero.
Of fifteen different graph topologies at the three-loop level
only the three graphs shown in fig. \ref{3loopgraphs}
contribute to this component.

Since those involve only the building block ``I'' which is already
known from the two-loop calculation, this immediately reduces
the original number of four integrations to the two integrals
over the insertion points. Moreover, it turns out that, after summing
up all terms, only two different integrals remain, namely the one-loop
box integral $h^{(1)}$ above and the two-loop integral
$h_{12}^{(2)}\equiv x_{12}^2 \int 
dx_5\int dx_6\, {1\over x_{15}^2x_{25}^2x_{35}^2x_{56}^2 
x_{16}^2x_{26}^2x_{46}^2}$, 
which is also known explicitly in
terms of polylogarithms up to fourth order \cite{10}.
Thus only a bit of algebra is required to arrive at the
following final
result for this
three-loop correlator \cite{11}:

\bear
G^{\rm 3-loop}
&=&-16{g^4\over (2\pi)^{16}}
{(N_c^2-1)N_c^2 R'\over x_{12}^2x_{23}^2x_{34}^2x_{41}^2} \Bigl[ 
(x_{12}^2x_{34}^2+x_{13}^2x_{24}^2+x_{14}^2x_{23}^2) (h^{(1)})^2 
\non\\
&&\hspace{110pt}
+4(h_{12}^{(2)}+h_{13}^{(2)}+h_{14}^{(2)}) 
\Bigr]
\non\\ \label{finalN=4} \ear\no 
This result has 
been confirmed by an $N=1$ superfield calculation in
\cite{12}. 

To summarize, 
we have used $N=2$ harmonic superspace for computing
the three-loop correlator of four supercurrents
in $N=4$, reaching a compact and explicit result
in terms of polylogarithms of the conformal cross
ratios. The computational effort in this calculation
was relatively small, due to a fortuitious interplay
between superconformal covariance arguments and Intriligator's
insertion formula. Considering the fact that the insertion
formula by itself seems almost a triviality at the path integral
level \cite{8}, its 
usefulness in the present context is quite
remarkable: Firstly, the fact that it involves {\sl chiral}
insertion points has allowed us to apply superconformal covariance
arguments in a way which would not have been possible for the
original, purely analytic amplitude, and led to a reduction in
the number of Feynman diagrams. Secondly, for the few remaining
diagrams we encountered, due to the factorization into simple
building blocks, only twofold integrals instead of the
fourfold integrals which had to be computed in the direct approach
of \cite{12}. Those twofold integrals were, moreover, {\sl individually}
conformal, which explains why only $h^{(1)}$ and
$h^{(2)}$ appeared; those are the only finite and conformally covariant 
integrals which one can build with the number of propagators 
available at the three-loop level.
Those two integrals are (as we indicated already by the superscript)
just the first two elements of the infinite series of conformal
``multi-ladder'' integrals $h^{(k)}$
defined and computed in
\cite{10}. 
This raises the interesting possibility that the $n$ - loop
contribution to this correlator may perhaps have the 
simple form $P_n(g^2h^{(1)},g^4h^{(2)},\ldots,g^{2n-2}h^{(n-1)})$
with some polynomial $P_n$.

We believe that even the corresponding four-loop calculation
would not be exceedingly difficult with the technique presented
here. Knowing the four-loop contribution would be interesting
also from the point of view of the Maldacena conjecture,
since for this correlator subleading color diagrams appear
first at this loop level.

%\noindent{\bf Acknowledgements.} 
%for sending their contributions before September 30, 2000.

\vspace{15pt}
{\bf \Large References}
\vspace{-10pt}
%%%%%%%%%%%%%%%%%%%%%%%%%%%%%%%%%%%%%%%%%%%%%%%%%%%%%%%%%%%%%%%%%%%%%%%%%%
%%%  The Bibliography is set using standard LaTeX macros

\end{document}